\documentclass[10pt,aps,prl,preprintnumbers,superscriptaddress,showpacs,
twocolumn,floatfix,nofootinbib]{revtex4-2}
\usepackage{graphicx}
\usepackage{amsmath}
\usepackage{slashed}

\newcommand{\bfm}[1]{\mbox{\boldmath$#1$}}

\newcommand{\gsim}{\;\rlap{\lower 3.5 pt \hbox{$\mathchar \sim$}} \raise 1pt
\hbox {$>$}\;}
\newcommand{\lsim}{\;\rlap{\lower 3.5 pt \hbox{$\mathchar \sim$}} \raise 1pt
\hbox {$<$}\;}

\begin{document}

\title{\boldmath
Oscillating Fields, Emergent Gravity and  Particle Traps
\unboldmath}
\author{Alexander A. Penin}
\email[]{penin@ualberta.ca}
\affiliation{Department of Physics, University of Alberta, Edmonton, Alberta T6G
2J1, Canada}
\author{Aneca Su}
\email[]{ssu2@ualberta.ca}
\affiliation{Department of Physics, University of Alberta, Edmonton, Alberta T6G
2J1, Canada}
\begin{abstract}
We study  the large-scale dynamics  of  charged particles  in a
rapidly oscillating  field and formulate its classical and
quantum effective theory description. The high-order
perturbative results for the effective action are presented.
Remarkably, the action models the effects of general relativity
on the motion of nonrelativistic particles, with the values of
the emergent curvature and speed of light determined by the
field spatial distribution and frequency. Our results can be
applied to a  wide range of physical problems including the
high-precision analysis and design of the charged particle
traps and Floquet quantum materials.
\end{abstract}

\maketitle

Since the classical work \cite{Kapitza:1951} the dynamics of
particles in a rapidly oscillating field has been studied in a
wide range of problems  from dynamical chaos
\cite{Friedman:2001} to quantum computing
\cite{Blatt:2008,Monroe:2021} and Floquet engineering of
quantum materials \cite{Oka:2019} with the renowned application
in the design of the Paul traps \cite{Paul:1990}.  Theoretical
description of this class of systems is based on the concept of
averaging, when the effect of the oscillating field is smeared
out and the long-time evolution is governed by the resulting
effective interaction naturally obtained within the {\it
high-frequency expansion} as a series in the ratio of the
oscillation period to a characteristic time scale of the
averaged system. The method is well known in classical
mechanics \cite{Bogoliubov:1961} and has been extended to
quantum systems
\cite{Cook:1985,Grozdanov:1988,Rahav:2003a,Rahav:2003b}. Many
subsequent works were dedicated to the quantum physics
applications and the method has been refined and generalized to
include many-body systems, spin, adiabatic variation of the
oscillating field {\it etc.}
\cite{Verdeny:2013,Goldman:2014,Eckardt:2015,Itin:2015,Mikami:2016,Bukov:2016,Weinberg:2017,Restrepo:2017}.
However, given the importance of the problem, surprisingly
little is know about the high-order perturbative behavior of
the generic three-dimensional systems  even at the classical
level. The existing analysis of the quantum systems based on
Floquet theory quickly becomes tedious in high orders too, and
often lacks the proper power counting. Hence, it is no surprise
that the theory of the charged particles confined  in the Paul
traps \cite{Leibfried:2003} is far less accurate than the one
for the Penning traps \cite{Brown:1985rh}. The goal of this work
is to introduce a new  foundation for a systematic analysis of
the periodically driven systems in the high-frequency limit.
Its core is the effective field theory approach  ideal for the
perturbative treatment of the  multiscale problems. We start
with the discussion of a classical system to identify the
relevant scales, expansion parameters, and  power counting
rules. Then we elaborate an asymptotic method to compute the
classical effective action  to high orders in high-frequency
expansion. Remarkably, the resulting effective interaction
models the dynamics of the nonrelativistic particle in the
pseudo-Riemann space, which gives a new nontrivial example of
``analog gravity'' \cite{Barcelo:2011}. To quantize the
effective action we develop the high-frequency effective theory
(HFET), being guided by an analogy  between the high-frequency
expansion and the nonrelativistic expansion of quantum
electrodynamics (QED).

Our  starting point is the classical equation of motion for a
particle of mass $m$ subjected to a static force $-\bfm
G$ and a periodic force $-{\bfm F}\cos \omega t$
\begin{equation}
\begin{split}
&m\ddot{\bfm R}+\bfm G(\bfm R)+\bfm F(\bfm R)\cos{\omega t}=0\,,
\label{eq::EOMfull}
\end{split}
\end{equation}
where the dot stands for the time derivative $d/dt$ and the
bold fonts indicate  three-dimensional vectors. The periodic
drive is limited to a single harmonic for the clarity of the
presentation but the inclusion of higher harmonics is rather
straightforward. We do not specify the nature of the external
fields to keep the discussion general and  consider  the limit
of fast oscillation. Let us quantify this condition as it plays
a crucial role for the determination of the expansion parameter
and the power counting rules. For a system of a characteristic
size $L$ the typical velocity acquired by the particle   under
the action of the time-independent force is ${v}\sim \left( {G}
L/m\right)^{1/2}$. One can define a ``reference'' velocity
$c=L\omega$ and the oscillations are considered fast when
$v/c\ll 1$. The main idea of the effective theory approach is
to separate the ``slow'' large-scale dynamics characterized by
the velocity $v$ from the ``fast'' small-scale  dynamics
characterized by the velocity $c$ and  manifested  through the
power corrections in the scale ratio to the effective action.
As we will see, the expansion in  $v/c$ shares many features
with the nonrelativistic expansion of the relativistic field
theories, with  $c$ playing a role of the speed of light.  It
is convenient to introduce the dimensionless variables $\omega
t\to t$, $R/L\to R$ so that the equation of motion becomes
\begin{equation}
\begin{split}
&\ddot{\bfm R}+\bfm g(\bfm R)+\bfm f(\bfm R)\cos{t} =0\,
\label{eq::EOMnd}
\end{split}
\end{equation}
with  $\bfm g=\bfm G/(Lm\omega^2)$ and $\bfm f=\bfm
F/(Lm\omega^2)$. Note that in the rescaled variables $c=1$ and
the expansion parameter is  $v$. While $\bfm  g=O(v^2)$ by
definition, the scaling of the oscillating term needs to be
determined. The leading contribution of the oscillating field
to the effective action is quadratic in its amplitude and we
are interested in the physical systems where the large-scale
dynamics is essentially determined  by the effect of the
periodic drive, which should be comparable to the one of the
static field.  This requires $\bfm  f=O(v)$, {\it i.e.} with
the rest of the parameters fixed, the amplitude of the
oscillating field should  scale linearly with its frequency.
This does not necessarily mean the actual dependence of the
amplitude on the frequency but rather determines the relevant
range for the ratio of  the static and oscillating field
magnitude at a given $\omega$. The above problem appears in a
variety of  physical systems and a number of  methods  have
been developed to disentangle the slow and fast dynamics in
perturbation theory. They share the principal  idea of
introducing independent variables for the fast and slow
evolution with subsequent averaging over the fast one. Its
particular realization, however, is crucial to get an efficient
tool for the high-order analysis. We follow the general idea of
the asymptotic method \cite{Bogoliubov:1961} and look for the
solution in the form
\begin{equation}
\begin{split}
&{\bfm R}={\bfm r}+\sum_{n=1}^\infty
\left[\bfm c_n({\bfm r})\cos(nt)+\bfm s_n({\bfm r})\sin(nt)\right]\,,
\label{eq::modes}
\end{split}
\end{equation}
where the vector ${\bfm r}$ describes the large-scale slow
evolution, $\dot{\bfm r}\equiv {\bfm v}=\sum_{m=1}^\infty {\bfm
v}^{(m)}({\bfm r})$ with ${\bfm v}^{(m)}=O(v^m)$. The method
\cite{Bogoliubov:1961} has been originally developed for the
nonlinear oscillation theory and its characteristic feature is
that the oscillation amplitude  itself is taken as a slow
variable. In the case of non-quasiperiodic motion at hand a
natural choice of the slow variable  is the path along the
smeared trajectory ${\bfm r}(t)$. Then the total time
derivative splits into the slow and fast components as follows
${d/dt}={\bfm v}\cdot\bfm\partial_r+\partial_t$. Substituting
Eq.~(\ref{eq::modes}) into Eq.~(\ref{eq::EOMnd}) and
reexpanding in the Fourier harmonics one can find the
coefficients $\bfm c_n({\bfm r})$ and $\bfm s_n({\bfm r})$
order by order in $v^2$. The zero harmonic then defines the
equation of motion for the slow evolution of the form
$\ddot{\bfm r}+{\bfm {\cal F}}_{\rm eff}({\bfm r},{\bfm v})=0$.
At ${\cal  O}(v^2)$ we get the well known  leading order
expression
\begin{equation}
\begin{split}
&{\bfm {\cal F}}_{\rm eff}={\bfm g}+{1\over 2}f_i\partial_i {\bfm f}\,.
\label{eq::EOMlo}
\end{split}
\end{equation}
The new next-to-leading ${\cal  O}(v^4)$ result reads
\begin{equation}
\begin{split}
&{\bfm {\cal F}}_{\rm eff}={\bfm g}+{1\over 2}f_i\partial_i {\bfm f}
-{3\over 2}v_iv_j\left(\partial_i\partial_jf_k\right)\partial_k{\bfm f}
+{1\over 4}f_if_j\partial_i\partial_j {\bfm g}\\
&
+\bigg[{3\over 2}g_i\left(\partial_if_k\right)
+{1\over 2}f_i(\partial_ig_k)+{25\over 32}f_i(\partial_if_j)
(\partial_jf_k)\\
&+{3\over 16}f_if_j(\partial_i\partial_jf_k)\bigg]\partial_k{\bfm f}
+{1\over 32}f_if_j(\partial_if_k)\partial_j\partial_k{\bfm f}\\
&+{1\over 16}f_if_jf_k\partial_i\partial_j\partial_k{\bfm f}\,,
\label{eq::EOMnlo}
\end{split}
\end{equation}
where the summation over repeating vector indices is implied.
So far we did not make any assumption about  the properties of
the fields. If we assume the existence of the corresponding
potentials $\bfm g={\bfm \partial}V_g$ and $\bfm f={\bfm
\partial}V_f$, Eqs.~(\ref{eq::EOMlo},\,\ref{eq::EOMnlo}) follow
from the effective Lagrangian
\begin{equation}
\begin{split}
&{\cal L}_{\rm eff}={{v_iv_j}\over 2}\left[\delta_{ij}
-{3\over 2}\partial_i{\bfm f}\partial_j{\bfm f}\right]
-V_{\rm eff}\,,
\label{eq::Leffnlo}
\end{split}
\end{equation}
where the effective potential reads
\begin{equation}
\begin{split}
&{V}_{\rm eff}=V_g+{{\bfm f}^2\over 4}
+{1\over 4}f_i{\bfm f}\partial_i{\bfm g}
+{1\over 64}f_if_j(\partial_i{\bfm f})\partial_j{\bfm f}\\
&+{1\over 16}
f_if_j{\bfm f}\partial_i\partial_j{\bfm f}\,.
\label{eq::Veffnlo}
\end{split}
\end{equation}
In the quadratic approximation in the oscillating field
the effective interaction has a distinctive form. The velocity
dependent term in Eq.~(\ref{eq::EOMnlo}) can be associated with
the geodesic equation for the affine connection
$\Gamma^k_{ij}=-{3\over 2}(\partial_i\partial_j{\bfm
f})\partial^k {\bfm f}$ corresponding to the three-dimensional
metric $\gamma_{ij}=\delta_{ij}-{3\over 2}\partial_i{\bfm
f}\partial_j{\bfm f}$. For an arbitrary field $\bfm f$ with
nonvanishing second derivative this metric describes a
non-Euclidean space.  In the region of vanishing charge density
${\bfm \partial}{\bfm f}=0$   the expression for the
corresponding Riemann  curvature scalar takes a  particulary
simple form $R^{(3)}={3\over 2}(\partial_i\partial_j{\bfm
f})^2$ and is non-negative.  Moreover, in the quadratic
approximation Eq.~(\ref{eq::Leffnlo}) coincides with the
post-Newtonian expansion of the relativistic Lagrangian for a
particle moving in a gravitational field ${\cal
L}=-(g_{\mu\nu}x^\mu x^\nu)^{1/2}$, where $x^\mu=(t,{\bfm r})$
and the metric of the $3+1$ dimensional pseudo-Riemann space
is\footnote{The product of the three-dimensional  vectors $\bfm
f$ is always defined with the Euclidean metric}
\begin{equation}
\begin{split}
&g_{00}=1+{\bfm f}^2/2,\quad g_{0i}=0,\quad g_{ij}=-\gamma_{ij}
\,.
\label{eq::metric}
\end{split}
\end{equation}
The corresponding scalar curvature reads
\begin{equation}
\begin{split}
&R^{(4)}=(\partial_i{\bfm f})^2
-{3\over 2}(\partial_i\partial_j{\bfm f})^2\,.
\label{eq::curvature}
\end{split}
\end{equation}
The above method readily generates  the higher order terms of
the effective Lagrangian and is  limited mainly by the size of
the resulting expressions. We present a relatively compact
${\cal O}(v^6)$ Lagrangian in one dimension  since many
physical systems can be reduced or decomposed  into the
one-dimensional problems. For a single generalized coordinate
$q$ we get the next-to-next-to-leading result
\begin{equation}
\begin{split}
&{\cal L}_{\rm eff}={\dot{q}^2\over 2}
\bigg[1-{3\over 2}f'^2-10f'f''g-10f'^2g'-3ff'g''\\
&-{379\over 128}f'^4-{691\over 64}ff'^2f''
-{3\over 128} f^2f''^2-{9\over 8}f^2f'f'''\bigg]\\
&+{\dot{q}^4\over 12}\bigg[-5f''^2+10f'f'''\bigg]
-\bigg[V_g+{f^2\over 4}+{f^2g'\over 4}+{f^2f'^2\over 64}\\
&+{f^3f''\over 16}-{5\over 4}f'^2g^2-{5\over 4}ff'^3g
+{f^2g'^2\over 4}+{9\over 256}f^2f'^2g'\\
&+{3\over 16}f^3f''g'+{f^3f'g''\over 64}+{f^4g'''\over 64}
-{1435\over 4608}f^2f'^4\\
&+{65\over 4608}f^3f'^2f''+{41\over 1152}f^4f''^2
+{f^4f'f'''\over 192}+{f^5f''''\over 384}
\bigg]\,,
\label{eq::Leffnnlo}
\end{split}
\end{equation}
where dash stands for the derivative $d/dq$.  Note that the
emergent Lorentz invariance of the effective action is broken
by the $\dot{q}^4$ term of Eq.~(\ref{eq::Leffnnlo}) in
agreement with the general argument \cite{Weinberg:1980kq}.

The result  Eq.~(\ref{eq::Leffnnlo}) has an interesting
connection to the theory of parametric resonance and stability
of dynamical systems, which is crucial for the further
discussion of the effective theory power counting. Namely, for
$g(q)=\delta q+{\cal O}(q^2)$ and $f(q)=\epsilon q+{\cal
O}(q^2)$ with some parameters $\delta$ and $\epsilon$ the
system has  an equilibrium point ${\cal F}=0$ at $q=\dot{q}=0$.
Then the equation $\partial {{\cal F}}(q,\dot{q})/\partial
q\big|_{q=\dot{q}=0} =0$ controls the change of its
stability.  This equation defines $\delta$ as a function of
$\epsilon$, {\it i.e.} one of the stability curves in the
parameter space which play a crucial role in the analysis of
chaotic and regular behavior of dynamical  systems. For
$\epsilon\ll 1$  through the next-to-next-to-leading
approximation we get
\begin{equation}
\begin{split}
& \delta=-{\epsilon^2\over 2}+{7\over 32}\epsilon^4
-{29\over 144}\epsilon^6+{\cal O}(\epsilon^8)\,,
\label{eq::stability}
\end{split}
\end{equation}
which is consistent  with the scaling $\bfm g\sim \bfm f^2$.
Eq.~(\ref{eq::stability}) agrees with the result obtained
within Floquet theory analysis of Mathieu equation
\cite{Kovacic:2018}, being a non-trivial test of our analysis.
This equation, in particular, defines the corrections to the
classical result on the stability of  inverted pendulum with
the natural frequency $\sqrt{-\delta}$ and the forced
oscillation amplitude $\epsilon$ \cite{Kapitza:1951}. Recently
the analysis of  the periodically driven pendulum with
$f,~g\propto \sin(q)$ has been performed to very high orders of
perturbation theory  \cite{Beneke:2023ndu}.
Eq.~(\ref{eq::Leffnnlo}) agrees with the next-to-leading
effective Lagrangian presented there. For higher orders the
comparison of the results is not straightforward since
in \cite{Beneke:2023ndu} the velocity dependent terms are
eliminated from  the equation of motion by using the energy
conservation. Hence, the resulting  effective potential depends
on  the total energy of the system, while we use the standard
definition of the Lagrangian  independent of the initial
conditions.

\begin{figure}[t]
\begin{center}
\includegraphics[width=1.5cm]{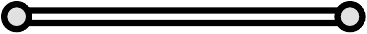}
\hspace*{2mm}\raisebox{0mm}{$\displaystyle{=}$}\hspace*{2mm}
\includegraphics[width=1.5cm]{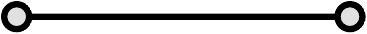}
\hspace*{2mm}\raisebox{0mm}{$\displaystyle{+}$}\hspace*{2mm}
\includegraphics[width=1.5cm]{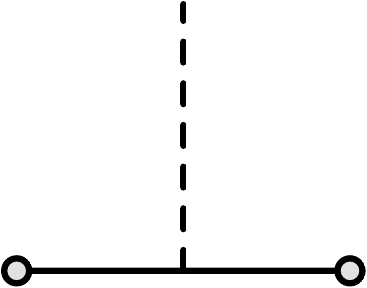}\\[7mm]
\hspace*{-1mm}\raisebox{4mm}{$\displaystyle{+}$}
\raisebox{4mm}{$\displaystyle{1\over 2!}\,\Bigg($}\hspace*{1mm}
\includegraphics[width=1.3cm]{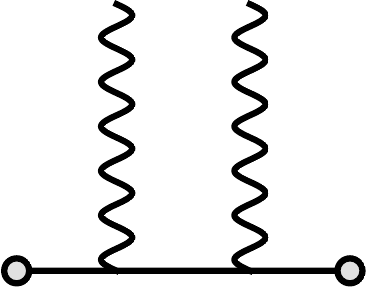}
\hspace*{2mm}\raisebox{4mm}{$\displaystyle{+}$}\hspace*{2mm}
\includegraphics[width=1.3cm]{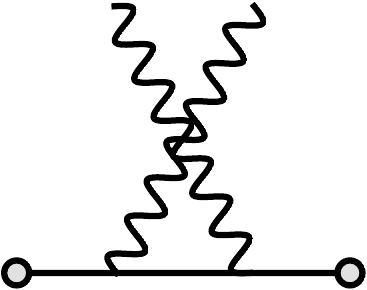}
\hspace*{1mm}\raisebox{4mm}{$\Bigg)$}
\hspace*{1mm}\raisebox{4mm}{$\displaystyle{+\hspace*{1mm}\ldots}$}
\end{center}
\caption{\label{fig::1} The Feynman diagrams representing the
expansion of the Green function Eq.~(\ref{eq::GFseries}). The
double (single) line represents the exact (free) particle
propagator while the dashed (wavy) line corresponds to the
static (oscillating) external field.}
\end{figure}

\begin{figure}[t]
\begin{center}
\includegraphics[width=1.7cm]{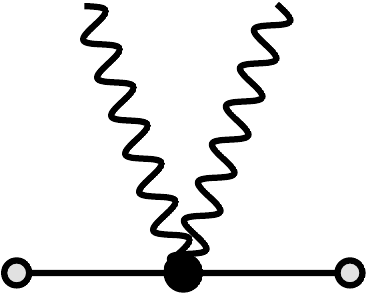}
\end{center}
\caption{\label{fig::2} The effective local vertex resulting
from the expansion Eq.~(\ref{eq::GF0series}) of the
off-shell propagator in Fig.~\ref{fig::1}.}
\end{figure}

Let us now consider the quantization of the effective action.
The existing theory of quantum systems in a rapidly oscillating
field is based on Floquet  analysis of Schr\"odinger  equation
with the time-periodic Hamiltonian ${\cal H}=\hat{\bfm
p}^2/2+V_g+V_f\cos t$, where $\hat{\bfm p}=-i\hbar\bfm\partial$
is the momentum operator and we keep the dependence on the
Planck constant $\hbar\ne 1$ to  separate the quantum
corrections from the classical action. The general idea of the
method is to construct within the high-frequency expansion a
unitary operator $\hat U$  such that the effective Hamiltonian
${\cal H}_{\rm eff}=\hat U^\dagger {\cal H}\hat U-i\hbar \hat
U^\dagger \partial_t{\hat U}$ is time-independent and
determines the quasienergy spectrum, {\it i.e.} the slow
evolution  of the quantum states. The particular realizations
of this program may be different. However, in this framework
the perturbative calculations quickly become tedious as the
order of approximation increases. At the same time the
multiscale problems are common to the quantum field theory,
where very efficient methods based on the scale separation are
elaborated and optimized for high-order calculations. As it was
pointed out, the high-frequency expansion  is  similar to the
nonrelativistic expansion and we suggest to realize  it in the
same way as the  Dirac equation in an external field is
expanded in inverse powers of the speed of light
\cite{Landau:1982}. Let us consider a Green function of the
original time-dependent Schr\"odinger  equation ${\cal
G}=(i\hbar\partial_t-{\cal H}+i\varepsilon)^{-1}$ and its
Fourier transform $\tilde {\cal G}(\bfm p_i,\bfm p_f;
{E_i},{E_f})$, which depends on the initial and final momentum
and energy variables. In general the initial and final energy
may differ due to the time dependence of the Hamiltonian.  We,
however, are interested in the low-energy behavior of the Green
function with the kinematical constraints\footnote{In this
section the explicit dependence on $\omega$, which is equal to
one in our system of units, is restored to indicate the
physical expansion parameter.} $\bfm p_{i,f}^2,~{E_{i,f}}\ll
\hbar \omega$. In this case  the periodic character of the time
dependence implies the energy conservation ${E_i}={E_f}\equiv
E$. Expanding the Green function in powers of the external
fields we get a series
\begin{equation}
\begin{split}
&\tilde {\cal G}=\tilde {\cal G}_0+
\tilde {\cal G}_0 \tilde V_g \tilde {\cal G}_0
+\tilde {\cal G}_0 \tilde  V_f \tilde {\cal G}_0
\tilde  V_f \tilde {\cal G}_0+\ldots\,,
\label{eq::GFseries}
\end{split}
\end{equation}
where $\tilde {\cal G}_0(\bfm p, {E})=({E}-\bfm
p^2/2+i\varepsilon)^{-1}$ is the free particle propagator and
$\tilde V_g$  ($\tilde V_f$) is the Fourier transform of $V_g$
($V_f\cos\omega t$).  The expansion is represented by the
Feynman diagrams in Fig.~\ref{fig::1}. Note that the
contribution with a single insertion of the oscillating  field
is forbidden by the ``energy scale'' conservation (for a
single-harmonic oscillating field this is true for any odd
power of $\bfm f$). Let us consider the  diagrams with the
double insertion of the oscillating field. The intermediate
state propagator carrying the momentum $\bfm p$ and energy ${
E}+\hbar\omega$ is far off-shell and can be expanded  in a
series
\begin{equation}
\begin{split}
\tilde {\cal G}_0(\bfm p, {E}+\hbar\omega)={1\over \hbar\omega}
-{{E}- \bfm p^2/2\over (\hbar\omega)^2}+\ldots\,,
\label{eq::GF0series}
\end{split}
\end{equation}
which gives rise to a local effective vertex, Fig.~\ref{fig::2}.
This  {\it seagull} vertex  is well know in  nonrelativistic
QED where it is generated by a far off-shell positron in the
intermediate state rather than  the large time-like momentum
transfer from the oscillating field. The odd powers  in $\omega
$ cancel between the planar and nonplanar  diagrams and by the
standard tools we readily get the leading  ${\cal
O}(1/\omega^2)$ contribution to the effective vertex in the
coordinate space
\begin{equation}
\begin{split}
&
{\langle V_f(\hat{\bfm p}^2/2-{E})V_f\rangle\over 2(\hbar\omega)^2}
={{\bfm f}^2\over 4\omega^2}\,,
\label{eq::vertexlo}
\end{split}
\end{equation}
where the matrix element is taken between on-shell states with
${\bfm p}^2/2=E$. For $\omega=1$ we recover the leading
contribution to the classical effective potential
Eq.~(\ref{eq::Veffnlo}).   At ${\cal
O}(1/\omega^4)$ the   contribution of the operator
$V_f(\hat{\bfm p}^2/2-{E})^3V_f$ to the effective vertex can be
computed in the same way  with the result
\begin{equation}
\begin{split}
&{1\over (2\omega)^4}\bigg[3\left(\{\hat{p}_i\hat{p}_j,
\partial_i{\bfm f}\partial_j{\bfm f}\}_+\!
+\!2\hat{p}_i\partial_i{\bfm f}\partial_j{\bfm f}\hat{p}_j
\right)\!+\hbar^2(\partial_i\partial_j{\bfm f})^2\bigg].
\label{eq::vertexnlo}
\end{split}
\end{equation}
The terms omitted in Eq.~(\ref{eq::GFseries}) give rise to the
effective vertices with the higher powers of the external
fields, which along with
Eqs.~(\ref{eq::vertexlo},\,\ref{eq::vertexnlo}) define the HFET
Feynman rules.  However, in the given order these vertices
reduce to the classical effective potential as in
Eq.~(\ref{eq::vertexlo}), {\it i.e.}  require no additional
calculation. Setting $\omega=1$ and switching back to the
velocity power counting  for the effective Hamiltonian trough
${\cal O}(v^4)$ we get
\begin{equation}
\begin{split}
&{\cal H}_{\rm eff}={1\over 8}\left[\{\hat{p}_i\hat{p}_j,\gamma^{ij}\}_+\!
+2\hat{p}_i\gamma^{ij}\hat{p}_j
\right]+{\hbar^2\over 16}(\partial_i\partial_j{\bfm f})^2+V_{\rm eff}\,,
\label{eq::Heffnlo}
\end{split}
\end{equation}
where $\gamma^{ij}=\delta_{ij}+{3\over 2}\partial_i{\bfm
f}\partial_j{\bfm f}+{\cal O}(v^4)$ is the inverse of the
metric tensor $\gamma_{ij}$ and  $V_{\rm eff}$ is given by
Eq.~(\ref{eq::Veffnlo}).

If  we assume  $\bfm\partial\bfm f=0$, Eq.~(\ref{eq::Heffnlo})
simplifies  to ${\cal H}_{\rm
eff}=\hat{p}_i\gamma^{ij}\hat{p}_j/2-{\hbar^2\over
12}R^{(3)}+V_{\rm eff}$. It has an interesting property  that
the  kinetic energy is not given by the covariant  Laplace
operator as required by the geometry of a genuine Riemann
space. Thus while classically the emergent nature of the metric
is revealed by the ${\cal O}(v^4)$ Lorenz symmetry violating
terms,  at  quantum level it is manifested already in the
leading kinetic energy operator sensitive to the short-distance
properties of the underlying fundamental theory.

The result Eq.~(\ref{eq::Heffnlo}) can be generalized to an
arbitrary number of harmonics in the periodically oscillating
field. The calculation of the  classical action in this case is
straightforward though the result is less elegant, and the
quantum corrections are given by the sum of
Eq.~(\ref{eq::vertexnlo}) over the harmonics weighted by the
(square of) the corresponding Fourier coefficients. As in the
nonrelativistic QED, the spin structure can be easily
incorporated in the Feynman rules of HFET. The quantization of
the theory through  ${\cal O}(v^6)$ does not pose a technical
challenge in the HFET  framework as well.

Let us now compare our approach to the high-frequency expansion
based on  Floquet theory. For the problem discussed in this
paper the  effective ${\cal O}(1/\omega^4)$ Hamiltonian in one
spatial dimension  has been derived for the first time in
\cite{Rahav:2003a} (a formal general expression  in a different
representation can be found in \cite{Mikami:2016}). This
analysis relies on a formal  power counting in $1/\omega$, with
both $\bfm g$ and $\bfm f$  treated as ${\cal O}(1/\omega^2)$
quantities. Hence the result does not account for   the terms
with the fourth power of $\bfm f$  present in
Eqs.~(\ref{eq::Veffnlo},\,\ref{eq::Heffnlo}).  However, this
power counting does not apply to the most  interesting physical
case of  dynamical stabilization realized {\it  e.g. } in the
Paul traps,  where the oscillating field results in a
qualitative change of the system behavior. The latter  requires
$\bfm g\sim \bfm f^2$ scaling, {\it cf.}
Eq.~(\ref{eq::stability}). In general, the Floquet theory
calculations in this order are already quite tedious even in
one dimension and without the more challenging ${\cal O}(\bfm
f^4)$ terms, while the quantization of the Hamiltonian within
HFET  requires only a ``one-line'' derivation of
Eq.~(\ref{eq::vertexnlo}).

To summarize, in this work we have presented a number of
results connecting  dynamical systems, general relativity and
quantum  theory. We have elaborated  an asymptotic method  to
systematically construct the effective action for particles
moving  in a rapidly oscillating field. The effect of the
oscillating field on the large-scale dynamics models the
pseudo-Riemann space of general relativity, with the curvature
determined by the field spatial distribution  and  the
effective value of the speed of light determined by the
oscillation frequency. While appearance of emergent gravity in
condensed matter systems has already  been predicted
\cite{Unruh:1981,Garay:1999sk} and observed experimentally (see
{\it e.g.} \cite{Kolobov:2019qfs}) for quasiparticle
propagation, the rapidly oscillating field creates the
gravity-like effect for  the classical charged particles.
Guided by the analogy with the nonrelativistic expansion of
QED, we have quantized the effective action and developed the
high-frequency effective theory, apparently the most powerful
analytic tool  for the perturbative analysis of the
periodically driven systems. It can be used in a wide range of
physical applications from  the high-precision analysis and
design of the charged particle traps to the Floquet engineering
of  quantum materials.

\vspace{2mm}
\noindent
{\bf Acknowledgments.}  The work of A.P. was supported in part
by NSERC and the Perimeter Institute for Theoretical Physics.
The work of A.S. is supported by NSERC.


\end{document}